\documentclass[a4paper,preprint,12pt]{elsarticle} 
\usepackage[english]{babel} 
\usepackage[latin1]{inputenc}
\usepackage{amssymb}  
\usepackage{amsmath}    
\usepackage{pb-diagram}     
\usepackage{epstopdf}                   
\usepackage{graphicx}        
 \usepackage{epsfig}    
\usepackage{color}
\usepackage{fancyhdr}     
 \usepackage{natbib}  
\usepackage{rotating}
\usepackage{pdflscape}
\usepackage[T1]{fontenc}        
\usepackage{url}                
\usepackage{mathtools}                 
\usepackage{vmargin}    
\usepackage{lineno}
\usepackage{setspace}
\usepackage{color,soul}
\usepackage[ruled,vlined]{algorithm2e}  
\usepackage[title]{appendix} 
\biboptions{square}   
\usepackage[colorlinks=true,linkcolor= blue,urlcolor=blue,citecolor=blue]{hyperref} 

\newcommand{\bo}[1]{\mathbf{#1}} 
\def\indic{\hbox{1\kern-.24em\hbox{I}}}      

\newcommand{\var}{\mathbb{V}}  
\newcommand{\esp}{\mathbb{E}}

\newcommand{\x}{x}
\newcommand{\X}{X}

\newcommand{\R}{\mathbb{R}}

\newcommand{\M}{f}

\newcommand{\T}{T}    
\newcommand{\N}{\mathbb{N}}      
\newcommand{\NN}{n}

\newcommand{\norme}[1]{\left|\left| #1 \right|\right|_{L^2}}

\newcommand{\normp}[1]{\left|\left| #1 \right|\right|_{p}}

\newtheorem{prop}{Proposition}{\bf}{\it} 
\newtheorem{defi}{Definition}{\bf}{\it}  

\newtheorem{theorem}{Theorem}{\bf}{\it}     
\newtheorem{lemma}{Lemma}{\bf}{\it}          
\newtheorem{rem}{Remark}{\bf}{\it} 
{\bf}{\it} 

\catcode`\"=\active 
\catcode`\"=\active 
\def"{\og\ignorespaces}
\def"{{\fg}} 
 
\usepackage{etoolbox}  

\makeatletter
\def\ps@pprintTitle{%
  \let\@oddhead\@empty
  \let\@evenhead\@empty
  \def\@oddfoot{\reset@font\hfil\thepage\hfil}
  \let\@evenfoot\@oddfoot
}
\makeatother

\begin{document}

\begin{frontmatter}     
\title{Kernel-based sensitivity indices for any model behavior and screening}
\author[a,b]{Matieyendou Lamboni\footnote{Corresponding author: matieyendou.lamboni[at]gmail.com/univ-guyane.fr, 03/11/2023}}        
\address[a]{University of Guyane, Department DFRST, 97346 Cayenne, French Guiana, France} 
\address[b]{228-UMR Espace-Dev, University of Guyane, University of R\'eunion, IRD, University of Montpellier, France.}    
                                    
\begin{abstract}
Complex models are often used to understand interactions and drivers of human-induced and/or natural phenomena. It is worth identifying the input variables that drive the model output(s) in a given domain and/or govern specific model behaviors such as contextual indicators based on socio-environmental models. Using the theory of multivariate weighted distributions to characterize specific model behaviors, we propose new measures of association between inputs and such behaviors. Our measures rely on sensitivity functionals (SFs) and kernel methods, including variance-based sensitivity analysis. The proposed $\ell_1$-based kernel indices account for interactions among inputs, higher-order moments of SFs, and their upper bounds are somehow equivalent to the Morris-type screening measures, including dependent elementary effects. Empirical kernel-based indices are derived, including their statistical properties for the computational issues, and numerical results are provided.   
\end{abstract}     
                                     
						  
\end{frontmatter}   

\section{Introduction}
Complex models or computer codes such as socio-environmental models (SEMs) are increasingly being developed and used to understand interactions and drivers that affect human well-being and the sustainability of the environment. Such information are used to support decision making for a wide range of problems and issues where the context may depend on the scales of the problem and the issues at stake.  SEMs can be very complex and non-linear, and uncertainty arises in many guises, such as in the representation of the system of interest and in the data available to study them. Sensitivity analysis (SA) is a valuable tool to quantify the strength of the various un-controllable (like climate and other exogenous factors) and controllable (like policy and management levers) drivers on contextual indicators of well-being and sustainability (\cite{elsawah20}).  
      
Contextual indicators of well-being and sustainability often represent specific system behaviors such as a sustainable system, a safe or critical domain.  Conceptually, specific model behaviors are transformations of complex models or systems. Specific model behaviors encountered in the literature are mainly twofold regarding their impacts on the initial distribution of the model inputs. Firstly, the class of objective functions (e.g., \cite{saltelli02,saltelli04}) consists of all transformations of the model outputs that do not modify the initial distribution of the model inputs. It includes transformations done by i) projecting the model outputs onto a given basis (\cite{campbell06,lamboni11,xiao18}), using the kernel-based principal components, using feature maps of the outputs (\cite{aronszajn50,scholkopf02,berlinet04}); ii) considering the probabilities of the stochastic and dynamic outputs to exceed a given threshold (\cite{lamboni14}), iii) using a regression-based classifier (\cite{lamboni16e}), and iv) considering the membership functions from either a crisp (a.k.a binary) or fuzzy clustering (\cite{roux21}). 
      
Secondly, we distinguish transformations that alter the initial distribution of the inputs such as considering the outputs values within a given cluster; constrained outputs or inputs; factor mapping (\cite{spear80,rose91,saltelli08,lamboni14}) or excursion sets (\cite{chevalier13,french13,fossum20}), that is, the input values leading to a response to exceed a given threshold; considering the model output values with different chances of being included in the analysis of interest. For instance, membership functions (MFs) from a crisp clustering is equal to one for the output values that comply with the model behavior of interest and zero otherwise, while those from a fuzzy clustering give mainly non-negative values to the output values according to the way that such values meet a criterion of interest. We can see that the crisp clustering MFs will lead to consider only a subset of the output values such as the sustainable system, safe or critical domain. 
  
Performing global SA (e.g., \cite{sobol93,borgonovo14,owen14b,gamboa14b,mara15,lamboni21,fort21}) on MFs allows for identifying the input variables that drive the model output(s) into a given domain of interest or a specific model behavior (\cite{lamboni14,roux21}). However, this analysis does not provide the drivers of specific model behaviors. Moreover, considering contextual indicators of well-being and sustainability for decision supporting may implicitly introduce new dependency structures among the factors and levers, and new interactions among such inputs. It may lead to rare events, where outputs do not follow the Gaussian distribution. 
The recent dependent multivariate SA (dMSA, \cite{lamboni21,lamboni21ar}) requires the distribution of the model inputs that complies with the model behavior of interest (i.e., the target inputs distribution) for performing SA. Since dMSA is well-suited for Gaussian distributed sensitivity functionals (SFs), it may be insufficient for exploring contextual indicators in presence of higher-order moments or heavy tailed distributions of SFs.  
                   
Our paper presents an approach that helps addressing such modeling issues in that it develops new and generic measures of association between specific model behaviors and the drivers or inputs of such models. Our approach accounts for i) dependencies among input factors and levers, ii) interactions among the input factors and levers even for non-independent factors, and iii) higher-order moments of SFs. The proposed measures rely on kernel methods (\cite{aronszajn50,berlinet04}) and the theory of multivariate weighted distributions (\cite{fisher34,rao65,patil78,mahfoud82,jain95,navarro06,barry17}). The well-known variance-based SA (\cite{sobol93,saltelli08,borgonovo14}) and multivariate SA (\cite{lamboni18a, lamboni21}), which account for only the second-order moments of SFs, are particular cases of our new measures. We also extend the Morris screening method (\cite{morris91}).  
			                          
The paper is organized as follows: after characterizing the model behavior of interest by i) identifying the appropriate weight function or the multivariate weighted distribution, and ii) deriving the cumulative distribution function (CDF) of the target inputs and its dependency model in Section \ref{sec:wdf}, we introduce the new measures of association between the target inputs and the model outputs in Section \ref{sec:kbeff}. We then provide the first-order and total kernel-based indices with the former index less than the latter. We also introduce a new upper bound of the total index. For the $\ell_1$-based kernel, the associated upper bound of the total index is equivalent to the Morris-type screening measures, including the dependent elementary effects (\cite{morris91,lamboni21}). Empirical kernel-based indices and their properties are provided in Section \ref{sec:est}. Section \ref{sec:test} provides illustrations for some model behaviors of interest such as the critical domain and a cluster of the output values. Section \ref{sec:con} concludes this work. 
              
						    				         	         	          	                        
\section*{General notations}              
For an integer $d \in\N\setminus\{0\}$, let $\bo{\X} := (X_1, \ldots, X_d )$ be a random vector having $F$ as the joint CDF and $F_j,\, j=1, \ldots, d$ as marginal CDFs. 
We use $F_j^{-1}$ for the inverse  of $F_j$, and $Y\stackrel{d}{=} Z$ when $Y, Z$ have the same CDF.\\   
For a non-empty subset $u \subseteq \{1, \ldots, d \}$, we use $|u|$ for its cardinality (i.e., the number of elements in $u$) and $(\sim u) := \{1, \ldots, d \}\setminus u$.  For a given $u$, we use $\bo{\X}_u := (\X_j, \, \forall\, j \in u)$ for a subset of inputs; $\bo{\X}_{\sim u} := (\X_j, \, \forall\,  j \in (\sim u))$ and we have the  partition $\bo{\X} = (\bo{\X}_u, \bo{\X}_{\sim u})$. \\ 
We use $\normp{\cdot}$ for the $p$-norm on $\R^\NN$, $\esp(\cdot)$ for the expectation, $\var(\cdot)$ for the variance,	$\xrightarrow{D}$ and $\xrightarrow{P}$ for the convergence in distribution and in probability, respectively.    
 
\section{Target inputs and outputs distributions}  \label{sec:wdf} 
This section aims at identifying the weight functions for  some specific model behaviors and determining the distribution of the target inputs, that is, the distribution of the inputs that complies with the model behaviors of interest.  
    
\subsection{Weight functions for some target outputs}\label{sec:swf}
\subsubsection{Case of multivariate response models}\label{sec:swf11}
For a vector-valued function $\M: \R^d \to \R^\NN$, the inputs $\bo{\X}$ have $F$ as the initial CDF (i.e., $\bo{\X} \sim F$), $\rho$ as the density, and $\M(\bo{\X})$ is the initial model outputs. It is clear that the outputs distribution is determined by the inputs distribution.\\ 
Namely, for a given domain of interest $D$, let us consider the outputs of interest (called the target outputs) given by 
$ 
\left\{ \begin{array}{l}    
 \M(\mathbf{X}) \quad \mbox{with} \;  \;  \bo{\X} \sim F \\
 \mbox{s.t.} \quad \M(\bo{X})  \in D\\  
\end{array} \right.
$. 
When the set $\left\{ \bo{x} \in \R^d : \M(\bo{x})  \in D \right\}$ is not empty, there exists a CDF (i.e., $F^w$) of the target inputs such that (see \cite{lamboni22b,lamboni21ar}) 
\begin{equation} \label{eq:tinp} 
\displaystyle 
 \left\{ \begin{array}{l}    
 \M(\mathbf{X}) \quad \mbox{with} \;  \; \bo{\X} \sim F \\
 \mbox{s.t.} \quad \M(\mathbf{X})  \in D\\  
\end{array} \right.  \stackrel{d}{=}  \M(\bo{X}^w),  \quad \mbox{with} \; 
\bo{X}^w \sim F^w \, .      
\end{equation}        
   
By considering the weight function $w(\bo{x})= \indic_{D}\left(\M(\bo{x}) \right)$ with $\indic_{D}(\bo{x})$ the indicator function, we can see  that the density function (i.e., $\rho^w$) of $\bo{\X}^w$ given by    
\begin{equation} \label{eq:wpdf}
\rho^w(\bo{x}) \propto  w(\bo{x}) \rho(\bo{x}) \, , 
\end{equation}   
allows for characterizing the associated target outputs or the behavior of interest. Such weight function is identical to the crisp or binary clustering of the initial model outputs. For any classifier $c : \R^\NN \to \R^p$ with $p \in \N\setminus \{0\}$, we can extend the above weight function as follows:   
$$   
w_1(\bo{x}) := \indic_{D}\left(c\circ \M(\bo{x}) \right) \, , 
$$    
where $c\circ \M$ stands for the composition of $c$ by $\M$. Thus, we can cover many clustering approaches (such as PCA, the kernel methods, random forest and logistic regression) by choosing the classifier $c$. To work with continuous weight functions, $\indic_{D}$ can be replaced with smooth functions such as the logistic function, and this leads to the second kind of weight functions, that is,   
$$
w_2(\bo{x}) := m \left( c \circ \M(\bo{x}) \right),\; \, \mbox{with}\; \,  m : \R^p \to \R_+ \, .
$$  
The membership functions from a fuzzy clustering (\cite{bezdek84,hoppner99}) are particular cases of $w_2$. The general expression of a weight function is given as follows: 
$$   
w_3(\bo{x}) := m \left( c \circ \M(\bo{x}) \right) \indic_{D}\left(m_2 \left( c_2 \circ \M(\bo{x}) \right) \right) \, ,   
$$ 
where  $c_2 : \R^\NN \to \R^p$ and $ m_2 : \R^p \to \R_+$ are given functions. The weight function $w_3$ can still be used for constraining, restricting and modifying directly the distribution of the initial inputs by taking $c \circ \M$ and $c_2 \circ \M$ as the identity functions, that is, $c \circ \M(\bo{x})=\bo{x}$. These types of weight functions directly or indirectly affect the initial distribution of the model inputs as well as the initial outputs distribution because the target inputs distribution $F^w$ induced the target outputs distribution and vice versa for deterministic functions or models. We then include them in a wide class of the outputs distributions of interest called the target outputs distributions. 
     
Since the weight functions guide the initial model outputs toward the behavior of interest, performing SA using the weight functions helps identifying the input variables that drive the initial outputs toward the target outputs defined by such weight functions (see e.g. \cite{lamboni14,roux21} for independent initial inputs). Such works can be extended to cope with weight functions including dependent and/or correlated input variables by using the dependent SA introduced in \cite{lamboni21,lamboni21ar}. 
                
\subsubsection{Extension to multivariate and functional outputs}\label{sec:swf12}
Consider $\Theta \subseteq \R$ and a function $\M: \R^d \times \Theta \to \R^\NN$ given by $\M(\bo{\X}, \theta) \in \R^\NN$ with $\theta \in \Theta$. The function $\M$ represents the multivariate and functional response model, including dynamic models. When $\Theta=\{\theta_0\}$, it becomes a vector-valued function. For such functions, one may consider the following weight functions
$$
w_5(\bo{x}) := L_\Theta \left( c \circ \M(\bo{x}, \theta), \beta \right) \indic_{D}\left(m_2 \left( c_2 \circ \M(\bo{x}), \theta \right),\, \forall\, \theta \in \Theta \right) \, ,     
$$
with $L_\Theta$ a desirability measure over all $\theta \in \Theta$ such as a loss function and $\beta$ some thresholds (see \cite{lamboni14} for more details). 
       
\subsection{Distribution of the target inputs} 
This section deals with the distribution and conditional distributions of continuous  variables following multivariate weighted distributions such as the target inputs. 
 
In the theory of multivariate weighted distributions (\cite{fisher34,rao65,patil78,mahfoud82,jain95,navarro06,barry17}), the density function of $\bo{\X}^w$ (i.e., $\rho^w$) given by Equation (\ref{eq:wpdf}) is known as the weighted density associated with the initial density $\rho$ and the weight function $w$. The weight function aims at altering the density of the initial inputs by assigning unequal chances to the initial observations of being included in the posterior analysis. The formal definition of weight functions and the weighted probability density function (PDF) are given below. We use $\esp_F$ for the expectation taking w.r.t. $F$ and $\bo{x} \in \R^d$. 

\begin{defi} (\cite{fisher34,rao65,patil78}) \label{def:wdist}   \\  
Let $w :\R^d \to \R_+$ be a non-negative function.  
   
$\quad$ (i)  When $\esp_F\left[w(\bo{\X}) \right] < \infty$, then $w$ is a weight function.
     
$\quad$ (ii) The weighted density (i.e., $\rho^w$) associated with $w$ and $\rho$ is given by 
\begin{equation} \label{eq:wdist}  
 \rho^w(\bo{x}) := \frac{w(\bo{x})}{\esp_F\left[w(\bo{\X}) \right]} \rho(\bo{x})  \, .
\end{equation}        
   
$\quad$ (iii) The random vector $\bo{\X}^w$ having $\rho^w$ as PDF and $F^w$ as CDF (i.e., $\bo{\X}^w \sim F^w$) is called  the weighted random vector.  
\end{defi} 
   
The marginal PDF of $\bo{\X}_{u}^w$ and the PDF of $\bo{\X}_{\sim u}^w$ conditional on $\bo{\X}_u^w$  (i.e., $\bo{\X}_{\sim u}^w | \bo{\X}_u^w$) are given by (\cite{mahfoud82,kocherlakota95,jain95,navarro06})  
\begin{equation} \label{eq:wmdf} 
\rho^w_u(\bo{\x}_{u}) :=  \frac{\esp_{F_{\sim u|u}}\left[w(\bo{\x}_u, \bo{\X}_{\sim u})\right]}{\esp_{F}\left[ w(\bo{\X}) \right]} \rho_u (\bo{\x}_u)\, ; 
\end{equation}      
\begin{equation} \label{eq:wcdf}    
\rho^w_{\sim u | u}(\bo{\x}_{\sim u}  | \bo{\x}_u) := \frac{w(\bo{\x})}{\esp_{F_{\sim u|u}}\left[w(\bo{\x}_u, \bo{\X}_{\sim u})\right]} \rho_{\sim u | u} (\bo{\x}_{\sim u}  | \bo{\x}_u) \, , 
\end{equation}            
where $F_u$ (resp. $F_{\sim u | u}$) denotes the CDF of $\bo{\X}_{u}$ (resp. $\bo{\X}_{\sim u} |\bo{\X}_{u}$), and $\rho_u$ (resp. $\rho_{\sim u | u}$) denotes the PDF of $\bo{\X}_{u}$ (resp. $\bo{\X}_{\sim u} | \bo{\X}_{u}$).  
Likewise,  we are going to use $F^w_u$ (resp. $F^w_{\sim u | u}$) for the distribution of $\bo{\X}_u^w$ (resp. $\bo{\X}_{\sim u}^w | \bo{\X}_u^w$).   

In what follows, we use $F_{ind} := \prod_{j=1}^d F_j$ and $F_{ind}(\bo{\x}) := \prod_{j=1}^d F_j(x_j)$. For independent initial variables, it is clear that  $F = F_{ind}$ or $\rho =\prod_{j=1}^d \rho_j$ with $\rho_j$ the PDF of $\X_j$. Thus,  
 $F^{w}_{ind}$ denotes the weighted distribution associated with $F_{ind}$ and $w$. 
                 
\begin{rem} \label{rem:mcpdfw} 
When $F = F_{ind}$, we have  
\begin{equation} \label{eq:wmdfin} 
\rho^w(\bo{x}) := \frac{w(\bo{x})}{\esp_F\left[w(\bo{\X}) \right]} \,  \prod_{j=1}^d \rho_j (x_j);
\qquad 
\rho^w_u(\bo{\x}_{u}) =   \frac{\esp_{F_{\sim u}}\left[w(\bo{\x}_u,\, \bo{\X}_{\sim u} )\right]}{\esp_{F}\left[ w(\bo{\X}) \right]} \prod_{j \in u} \rho_j(x_j) \, ;  \nonumber
\end{equation}                    
\begin{equation} \label{eq:wcdfin} 
\rho^w_{\sim u | u}(\bo{\x}_{\sim u}  | \bo{\x}_u) =  \frac{w(\bo{\x})}{\esp_{F_{\sim u}}\left[w(\bo{\x}_u,\, \bo{\X}_{\sim u} )\right]} \prod_{j \in \{1, \ldots, d\}\setminus u} \rho_j(x_j) \, .  \nonumber 
\end{equation}               
\end{rem}  

Since the expressions of the marginal and conditional PDFs in Remark \ref{rem:mcpdfw} are much easier to work with, it is worth expressing any weighted random vector $\bo{\X}^w$ as the random vector associated with $\bo{Y} \sim F_{ind}$ and a new weight function (i.e., $w_e$) (see Proposition \ref{prop:neww}). To that end, we use  $C : [0,\, 1]^d \to [0,\, 1]$ for the copula of the initial inputs $\bo{\X}$, that is, $F(\bo{\x}) = C(F_1(x_1), \ldots, F_d(x_d))$ (\cite{sklar59,nelsen06,joe14}). 
       
\begin{prop} \label{prop:neww}   
Let $\bo{Y} \sim F_{ind}$ and assume that the copula $C$ has $c(\cdot)$ as the joint PDF. Then, the PDF of $\bo{\X}^w$ is given by 
$$
\rho^w(\bo{x})  = \frac{w_e(\bo{x})}{\esp_{F_{ind}} [w_e(\bo{Y})]} \prod_{j=1}^d  \rho_j(x_j);
\qquad
w_e(\bo{x}) := w(\bo{x}) c\left(F_1(x_1), \ldots, F_{d}(x_{d})\right)
 \, .   
$$
\end{prop}   
\begin{preuve}
See Appendix \ref{app:prop:neww}.  
\begin{flushright}  
$\Box$  
\end{flushright} 
\end{preuve}  
  
It comes out from Proposition \ref{prop:neww} that the copula-based expressions of the PDF of $\bo{\X}$ allow for writing 
\begin{itemize}
\item $F= F_{ind}^{c}$ to say that the initial distribution $F$ is an adjustment of the distribution $F_{ind}$ using the weight function $c$; 
\item $F^w= F_{ind}^{w_e}$  to say $F^w$ is the weighted distribution associated with $F_{ind}$  and $w_e$.   
\end{itemize} 
 
To deduce the CDF of the random vector $\bo{\X}^w \sim F^w=F_{ind}^{w_e}$, including its copula (see Theorem \ref{theo:copbcd}), we use $\boldsymbol{\pi} :=(\pi_1, \ldots, \pi_{|\boldsymbol{\pi}|})$ for an arbitrary permutation of $(\sim u) :=\{1, \ldots, d\}\setminus u$ with the cardinality $|\boldsymbol{\pi}|= d-|u|$. We also use $\bo{u}_{ \boldsymbol{\pi}} :=(u_{\pi_1}, \ldots, u_{\pi_{|\boldsymbol{\pi}|}}) \in [0, 1]^{d-|u|}$; $\bo{V}_{\boldsymbol{\pi}} :=\left(V_{\pi_k} \sim \mathcal{U}(0,\, u_{\pi_k}), \, k=1, \ldots, |\boldsymbol{\pi}| \right)$ for a random vector of independent variables. 
\begin{theorem} \label{theo:copbcd} 
Consider the random vector $\bo{\X}^w \sim F^w =F_{ind}^{w_e}$ and $\bo{Y} \sim F_{ind}$. Assume that $\bo{\X}^w$ and $\bo{Y}$ are continuous random vectors. 
Then,  there exist a CDF $W(\cdot ;\bo{x}_{u}^w) : [0, \, 1]^{d-|u|} \to [0, 1]$ given by 
\begin{equation}     \label{eq:wcopbcdf}
W\left(\bo{u}_{|\boldsymbol{\pi}|}; \bo{x}_{u}^w \right) =\frac{\esp_{\bo{V}_{\boldsymbol{\pi}}} \left[w_{e}\left(\bo{x}^w_u, F_{\pi_1}^{-1}(V_{\pi_1}), \ldots, F_{\pi_{|\boldsymbol{\pi}|}}^{-1}(V_{\pi_{|\boldsymbol{\pi}|}}) \right) \right]}{\esp_{F_{ind}}\left[w_{e}(\bo{x}^w_u, \bo{Y}_{\sim u}) \right]}\,  \prod_{j=1}^{d-|u|} u_{\pi_j} \, , 
\end{equation}          
such that 
\begin{equation}     \label{eq:wcdfw}
F_{\sim u|u}^{w}(\bo{x}_{\sim u} | \bo{x}_{u}^w) = W\left(F_{\pi_1}(x_{\pi_1}), \ldots, F_{\pi_{|\boldsymbol{\pi}|}}(x_{\pi_{|\boldsymbol{\pi}|}}); \bo{x}_{u}^w \right)
 \, .
\end{equation}     
\end{theorem}  
\begin{preuve}       
See Appendix \ref{app:theo:copbcd}.  
\begin{flushright}   
$\Box$ 
\end{flushright}
\end{preuve} 

It is worth noting that $W(\cdot ;\bo{x}_{u}^w)$ is a CDF of a random vector having $(0, \, 1)^{d-|u|}$ as the joint support, and it involves only the weight function and the marginal CDFs of $\bo{\X}$. When $u=\emptyset$, we obtain the following expressions of $W$:  
\begin{equation}     \label{eq:wcopbdf} 
W\left(u_{1}, \ldots, u_{d} \right) 
= \frac{\esp_{\bo{V}} \left[w_{e}\left(F_{1}^{-1}(V_{1}), \ldots, F_{d}^{-1}(V_{d}) \right) \right]}{\esp_{F_{ind}}\left[w_{e}(\bo{Y}) \right]} \prod_{k=1}^{d} u_{k} \, ,
\end{equation}    
and  
\begin{equation}     \label{eq:wcdfwful}
F^{w}(\bo{x}) = W\left(F_{1}(x_{1}), \ldots, F_{d}(x_{d}) \right)  \, .      
\end{equation} 
  
Equations (\ref{eq:wcopbcdf})-(\ref{eq:wcdfwful}) are useful for sampling random values of  $\bo{\X}_{\sim u}^w | \bo{\X}_{u}^w$ and $\bo{\X}^w$, and for deriving the dependency models of $\bo{\X}^w$.  
	                                  
\section{Effects of the target inputs on the target outputs}\label{sec:kbeff}
Since introducing weight functions in our analysis often leads to dependent target inputs, and the target inputs can lead to rare events, this section aims at proposing and studying kernel-based sensitivity indices for identifying the input variables that govern the target outputs. 
                                                        
\subsection{Sensitivity functionals based on weight functions} \label{sec:sfs}  
Sensitivity functionals (SFs) contain the primary information about the contribution of the model inputs (see \cite{lamboni16b,lamboni18,lamboni19,lamboni18a,lamboni22} for independent input variables, and \cite{lamboni21,lamboni21ar,lamboni22mcap} for dependent and/or correlated input variables). The first-order SF allows for quantifying the single contribution of the inputs, while the total SF is used to measure the overall contribution of the inputs, including interactions.    
              
For dependent weighted random vector (i.e., $\bo{\X}^w \sim F^w$) having the CDFs given by Equations (\ref{eq:wcopbcdf}), (\ref{eq:wcdfw}) and (\ref{eq:wcopbdf}), 
the dependency models of $\bo{\X}^w$ (\cite{skorohod76,lamboni21,lamboni21ar,lamboni22b,lamboni22mcap}) are given by  
\begin{equation} \label{eq:wdep}  
\bo{\X}^w_{\sim u} = r\left(\bo{\X}_{u}^w, \bo{U}_{\boldsymbol{\pi}} \right) \stackrel{d}{=} \left(
 F_{\pi_{1}}^{-1} \left( Z_{\pi_1} \right),
  \ldots,
F_{\pi_{|\boldsymbol{\pi}|}}^{-1} \left( Z_{\pi_{|\boldsymbol{\pi}|}} \right)  \right)  \, ,     
\end{equation}   
where $r$ is a vector-valued function;  $\bo{U}_{\boldsymbol{\pi}} :=\left(U_{\pi_1}, \ldots, U_{\pi_{|\boldsymbol{\pi}|}} \right) \sim \mathcal{U}(0,\, 1)^{d-|u|}$ and  
\begin{equation} \label{eq:wdepu}   
 \left[     
\begin{array}{l}      
Z_{\pi_1} :=   W_{\pi_1}^{-1}\left(U_{\pi_1} \, |  \bo{\X}_{u}^w \right) \\
 Z_{\pi_2} := W_{\pi_2 | \pi_1}^{-1}\left(U_{\pi_2} \, |  \bo{\X}_{u}^w, Z_{\pi_1} \right) \\  
 \vdots \\   
Z_{\pi_{|\boldsymbol{\pi}|}} := W_{\pi_{|\boldsymbol{\pi}|} |\sim \pi_{|\boldsymbol{\pi}|}}^{-1}\left(U_{\pi_{|\boldsymbol{\pi}|}} \, |\, \bo{\X}_{u}^w,\, Z_{\pi_1},\ldots, Z_{\pi_{|\boldsymbol{\pi}| -1}} \right)   \\     
							\end{array}  \right] \sim  W\left(\cdot \, |  \bo{\X}_{u}^w \right)  \, , \nonumber  
\end{equation}    
with $W_{\pi_2 |\pi_1}^{-1}$ the inverse of the CDF of $Z_{\pi_2}$ conditional on $Z_{\pi_1}$. Composing the target outputs $\M(\bo{\X}^w,\, \theta)$ by the dependency model given by (\ref{eq:wdep}) yields
$$ 
\M(\bo{\X}^w,\, \theta)  \stackrel{d}{=} \M\left(\bo{\X}^w_u,  r\left(\bo{\X}_{u}^w, \bo{U}_{\boldsymbol{\pi}} \right), \theta\right) \, .  
$$      
The first-order and total SFs associated with $\bo{\X}^w_u$ are respectively defined by   
\begin{eqnarray} 
\M_u^{fo}(\bo{\X}_u^w, \theta)  & \stackrel{}{:=} & \esp_{U} \left[\M\left(\bo{\X}^w_u,  r\left(\bo{\X}_{u}^w, \bo{U}_{\boldsymbol{\pi}} \right), \theta\right)\right] - \esp\left[\M\left(\bo{\X}^w_u,  r\left(\bo{\X}_{u}^w, \bo{U}_{\boldsymbol{\pi}} \right), \theta\right)\right] \, , \nonumber  
\end{eqnarray}   
$$          
\M_u^{tot}(\bo{\X}^w, \bo{U}_{\boldsymbol{\pi}},  \theta)  := \M\left(\bo{\X}^w_u,  r\left(\bo{\X}_{u}^w, \bo{U}_{\boldsymbol{\pi}} \right), \theta\right) -  \esp_{\bo{\X}^w_u} \left[\M\left(\bo{\X}^w_u,  r\left(\bo{\X}_{u}^w, \bo{U}_{\boldsymbol{\pi}} \right), \theta\right)\right] \, . 
$$     
Using the fact that $F^w = F^{w_e}_{ind}$,  new expressions of SFs based on $w_e$ are derived in Proposition \ref{prop:tef}.  To that end, we use $F_U$ for the CDF of $\mathcal{U}(0,\, 1)^{d-|u|}$ in what follows. 
\begin{prop}  \label{prop:tef}    
Let $\bo{\X}^w \sim F^w = F_{ind}^{w_e}$, $\bo{Y} \sim F_{ind}$ and $\bo{U} \sim F_U$ be independent random vectors. Then, the first-order and total SFs related to $\bo{\X}_u^w$ are given by   
\begin{equation} \label{eq:fosfw2} 
\M_u^{fo}(\bo{\X}_u^w, \theta)  =  \esp_{U} \left[\M\left(\bo{\X}^w_u,  r\left(\bo{\X}_{u}^w, \bo{U} \right),\, \theta\right)\right] -  
\frac{\esp\left[\M\left(\bo{Y}_u,  r\left(\bo{Y}_{u}, \bo{U} \right),\, \theta\right) w_e\left(\bo{Y}\right)\right]}{\esp\left[w_e\left(\bo{Y}\right)\right]} \, ; 
\end{equation}           
\begin{equation} \label{eq:tesfw} 
\M_u^{tot}(\bo{\X}_u^w, \bo{U}, \theta)  = \M(\bo{\X}_u^w,  r\left(\bo{\X}_u^w, \bo{U} \right), \theta) - \frac{\esp_{\bo{Y}} \left[\M(\bo{Y}_u, r_u\left(\bo{Y}_u, \bo{U} \right), \theta) w_e\left(\bo{Y}\right)\right]}{\esp \left[w_e\left(\bo{Y}\right)\right]} \, . 
\end{equation} 
\end{prop}  
\begin{preuve}  
The results are straightforward bearing in mind Remark \ref{rem:mcpdfw} (see also \cite{lamboni22mcap}). 
\begin{flushright} 
$\Box$ 
\end{flushright} 
\end{preuve} 

It is worth noting that the above SFs can be adapted for structured model outputs such as categorical variables, sequences, graphs and labels. 
   
\subsection{Definition and properties of the kernel-based  sensitivity indices} \label{sec:kbsa}
For the set of output values $\mathcal{Y}$ and  $\bo{y}, \bo{y}' \in \mathcal{Y}$, it is realistic i) to map a class of labels or contentious variables $\mathcal{Y}$ into a feature space $\mathcal{F}_k \subseteq \R^L$ using the map $\bo{y} \mapsto  \phi(\bo{y}) := \left\{\psi_\ell(\bo{y}) \right\}_{\ell=1}^L$ with $\psi_\ell$ a function and the possibility $L=\infty$; and ii) to describe the desirable structure by a similarity measure between $\bo{y}, \bo{y}'$ a.k.a. kernel, that is,     
$$    
k\left(\bo{y},\, \bo{y}'\right) : = \left<\phi(\bo{y}),  \phi(\bo{y}')\right>_{\mathcal{F}_k} \, ,  
$$  
where $\left< \cdot, \cdot \right>_{\mathcal{F}_k} $ stands for the inner product. This kernel is symmetric and positive definite (SPD), and the Moore-Aronszajn theorem (\cite{aronszajn50,berlinet04}) ensures that the feature space $\mathcal{F}_k$ is a RKHS endowed with the inner product $k\left(\cdot,\, \cdot \right)$. The feature map allows not only for dealing with the structured outputs but also for considering desirable moments. In what follows, assume that 
     
(A1): $k\left(\bo{y},\, \bo{y}'\right)$ is SPD; $\phi(\bo{y}) = k\left(\bo{y},\, \cdot\right)$ is convex and $k\left(\bo{y},\, \bo{0}\right) =0$.   
  
\noindent      
\textbf{Example:} the $\ell_p$-based kernel and the quadratic kernel given by
$$    
k_p\left(\bo{y},\, \bo{y}'\right) : = \left< \normp{\bo{y}}^p,  \normp{\bo{y}'}^p\right>; 
\qquad \quad   
k_q(\bo{Y}, \bo{Y}') := \left< \bo{Y}, \bo{Y}' \right>^2 \, ,
$$ 
respectively, satisfy (A1). 
  
Namely, we use $\bo{\X}^{w '}$ (resp. $\bo{U}'$) for an i.i.d. copy of $\bo{\X}^{w} \sim F_{ind}^{w_e}$ (resp. $\bo{U} \sim F_U$), and we consider the following functionals
$$
\M^c(\bo{\X}^w_u, \bo{U}, \theta) := \M(\bo{\X}_u^w,  r\left(\bo{\X}_u^w, \bo{U} \right), \theta) - \esp \left[\M(\bo{\X}_u^w,  r\left(\bo{\X}_u^w, \bo{U} \right), \theta) \right] \, ;   
$$
$$
\M_u^{*}\left(\bo{\X}^w_u, \bo{\X}^{w'}_u, \bo{U}, \theta\right) := \M\left(\bo{\X}^w_u,  r\left(\bo{\X}_{u}^w, \bo{U} \right), \theta\right) - \M\left(\bo{\X}^{w'}_u,  r\left(\bo{\X}^{w'}_u, \bo{U} \right), \theta\right) \nonumber \, .
$$    
For a mensurable and SPD kernel $k$, we also assume that  
                 
(A2): $ 0< \int_{\Theta} \esp\left[k\left( \M^c(\bo{\X}^w_u, \bo{U}, \theta),\, \M^c(\bo{\X}_u^{w '}, \bo{U}', \theta) \right)\right]\, d\theta < +\infty$. 
    
\subsubsection{Kernel-based  sensitivity indices for vector-valued functions}
This section deals with a vector-valued function $\M: \R^d \to \R^\NN$, which is a particular case of $\M(\bo{\X}, \theta)$, that is, when $\theta \in \Theta =\{\theta_0\}$. Thus, the above first-order and total SFs, the centered outputs $\M_u^{c}$ and $\M_u^{*}$ become respectively    
$$  
\M_u^{fo}(\bo{\X}^w_u);   
\quad \quad
\M_u^{tot}(\bo{\X}^w_u, \bo{U});
\quad \quad
\M^c(\bo{\X}^w_u, \bo{U});  
\quad \quad
\M_u^{*}\left(\bo{\X}^w_u, \bo{\X}^{w'}_u, \bo{U}\right) \, .  
$$       
We also use    
\begin{equation}  
\mu_u^{* ^p} :=  \esp\left[ \normp{\M_u^{*}\left(\bo{\X}^w_u, \bo{\X}^{w'}_u, \bo{U} \right)}^p\right]  \, ,    
\end{equation} 
 and we are going to see latter that the functional $ \M_u^{*}(\bo{\X}^w_u, \bo{\X}^{w'}_u, \bo{U})$ will lead to the upper bounds of the kernel-based SIs (Kb-SIs). Definition \ref{def:kbsi} formally introduces such indices. 
       
\begin{defi} \label{def:kbsi}
Let $\bo{\X}^{w '}$, $\bo{\X}^{w ''}$, and $\bo{\X}^{w '''}$ be i.i.d. copies of $\bo{\X}^{w}$, and assume that (A1)-(A2) hold. Then, the first-order and total Kb-SIs of $\bo{\X}^w_u$ are defined as follows: 
\begin{equation}
S_u^{k} := \frac{\esp\left[k\left(\M_u^{fo}(\bo{\X}^w_u),\, \M_u^{fo}(\bo{\X}^{w '}_u) \right) \right]}{\esp\left[k\left(\M^c(\bo{\X}^w_u, \bo{U}),\, \M^c(\bo{\X}^{w '}_u, \bo{U}') \right)\right]} \, ;  
\end{equation}    
\begin{equation}
S_{T_u}^{k} :=  \frac{\esp\left[k\left(\M_u^{tot}(\bo{\X}^w_u, \bo{U}),\, \M_u^{tot}(\bo{\X}^{w '}_u, \bo{U}') \right) \right]}{\esp\left[k\left(\M^c(\bo{\X}^w_u, \bo{U}),\, \M^c(\bo{\X}^{w '}_u, \bo{U}') \right)\right]} \, ,
\end{equation} 
respectively. We also define     
\begin{equation}  \label{eq:upksi}
\Upsilon_u^k :=\frac{\esp\left[k\left(\M^*_u(\bo{\X}^w_u, \bo{\X}^{w '}_u,\bo{U}),\, \M^*_u(\bo{\X}^{w ''}_u, \bo{\X}^{w '''}_u, \bo{U}') \right)\right]}{\esp\left[k\left(\M^c(\bo{\X}^w_u, \bo{U}),\, \M^c(\bo{\X}^{w '}_u, \bo{U}') \right)\right]} \,  . 
\end{equation}   
\end{defi}    
For the $\ell_p$-based kernel, we can see that 
$
\Upsilon_u^{k_p} =\frac{\left(\mu_u^{* ^p}\right)^2}{\esp\left[k_p\left(\M^c(\bo{\X}^w_u, \bo{U}),\, \M^c(\bo{\X}^{w '}_u, \bo{U}') \right)\right]}
$.       \\  
    
Lemma \ref{lem:pro} guarantees the usual and interesting properties of sensitivity indices one should expect,  such as taking into account the interactions among the target inputs does not reduce the effects of the associated inputs. It also provides the upper bound of the total index of $\bo{\X}_u^w$. 
 
\begin{lemma} \label{lem:pro} 
Consider a mensurable and SPD kernel $k$ and assume that (A1)-(A2) hold. Then, we have 
\begin{equation} 
0\leq  S_u^k \leq S_{T_u}^k \leq 1 \, ,     
\end{equation}    
\begin{equation}
 S_{T_u}^k \leq  \Upsilon_u^k \, . 
\end{equation}   
\end{lemma}     
\begin{preuve} 
See Appendix \ref{app:lem:pro}. 
\begin{flushright} 
$\Box$
\end{flushright}
\end{preuve}   
  
It comes out from Lemma \ref{lem:pro} that the upper bound of the total index given by (\ref{eq:upksi}) does not use the model derivatives. It is also worth noting that the numerator of the upper bound $\Upsilon_u^k$ relies on the difference of two model evaluations (i.e., $\M^*_u$), while the numerator of the total index requires the evaluation of the conditional expectation of the target outputs. As a matter of fact,  the computation of $\Upsilon_u^k$ should require less model evaluations to converge compared to the total index in general. Thus,  $\Upsilon_u^k$ can be used for screening the target inputs. Furthermore, we can significantly reduce the number of model evaluations for computing $\Upsilon_u^k$ by making use of the Morris sequence (\cite{morris91}) and the transformation $\X_j =F_j^{-1}(U_j)$. Since the kernel-based SIs include Sobol' indices, generalized sensitivity indices (GSIs) and dependent GSIs (see Section \ref{sec:link}), $\Upsilon_u^k$ is also the upper bounds of such indices.  
       
\subsubsection{Extension to multivariate and functional outputs}
To deal with spatio-temporel model and dynamic models, we extend the kernel-based SIs from Definition \ref{def:kbsi} as follows: 
\begin{equation} 
S_u^{k, f} := \frac{\int_\Theta  \esp\left[k\left(\M_u^{fo}(\bo{\X}^w_u, \theta),\, \M_u^{fo}(\bo{\X}^{w '}_u, \theta) \right) \right] \, d\theta}{ \int_\Theta\esp\left[k\left(\M^c(\bo{\X}^w_u, \bo{U}, \theta),\, \M^c(\bo{\X}^{w '}_u, \bo{U}', \theta) \right)\right] \, d\theta} \, ;  
\end{equation}     
\begin{equation}
S_{T_u}^{k, f} :=  \frac{\int_\Theta \esp\left[k\left(\M_u^{to}(\bo{\X}^w, \bo{U},  \theta),\, \M_u^{tot}(\bo{\X}^{w '}_u, \bo{U}', \theta) \right) \right] \, d\theta}{ \int_\Theta\esp\left[k\left(\M^c(\bo{\X}^w_u, \bo{U}, \theta),\, \M^c(\bo{\X}^{w '}_u, \bo{U}', \theta) \right)\right] \, d\theta} \, ;  
\end{equation} 
\begin{equation}
\Upsilon_u^{k,f} :=\frac{\int_\Theta \esp\left[k\left(\M^*_u(\bo{\X}^w_u, \bo{\X}^{w '}_u,\bo{U}, \theta),\, \M^*_u(\bo{\X}^{w ''}_u, \bo{\X}^{w '''}_u, \bo{U}', \theta) \right)\right] \, d\theta}{ \int_\Theta\esp\left[k\left(\M^c(\bo{\X}^w_u, \bo{U}, \theta),\, \M^c(\bo{\X}^{w '}_u, \bo{U}', \theta) \right)\right] \, d\theta} \,  . 
\end{equation}       

Such indices also satisfy the usual and interesting properties provided in Lemma \ref{lem:pro}. Moreover, when $\Theta=\{\theta_0\}$, we can remark that $S_u^{k, f}$,  $S_{T_u}^{k, f}$, and  $\Upsilon_u^{k,f}$ come down to $S_u^{k}$,  $S_{T_u}^{k}$, $\Upsilon_u^{k}$, respectively.
         
\subsection{Links with existing global sensitivity indices} \label{sec:link}
\subsubsection{On the $\ell_1$-based kernel indices and Morris-type screening methods} 
When $\NN =1$, $k_1\left(y, y' \right) := |y||y'|$, the numerator of $\Upsilon_j^{k_1}$ for any $j \in \{1, \ldots, d\}$ is given by 
$$
\left(\mu_j^{* ^1} \right)^2 :=  \esp^2 \left[\left| \M_j^*\left(\X^w_j, \X^{w'}_j, \bo{U}\right)\right| \right] =   \esp^2 \left[\left| \M\left(\X^w_j, r\left(\X^w_j, \bo{U} \right)\right)-\M\left(\X^{w'}_j, r\left(\X^{w'}_j
 \bo{U}\right) \right) \right| \right] \, .      
$$   
Using the Morris sequence, we can see that the empirical expression of $\mu_j^{* ^1} = \esp \left[\left| \M_j^*\left(\X^w_j, \X^{w'}_j, \bo{U}\right)\right| \right]$ is proportional to the Morris-type screening measures, that is, $\mu^*_j$ for independent variables and its extension $d\!\mu_j^*$ provided in \cite{lamboni21} for dependent inputs.  Thus, the Morris-type screening measures $\mu^*_j$ and $d\!\mu_j^*$ are the upper bounds of $\sqrt{S_{T_j}^{k_1}}$ up to a constant of normalization. The normalized upper bound $\Upsilon_j^{k_1}$ enhances such screening measures. As a consequence, the Morris-type screening measures implicitly require working with the $\ell_1$-norm or applying the $\ell_1$-norm on SFs. Moreover, as small norms allow for capturing slow variations, the new $\ell_1$-based kernel SIs and the Morris-type screening measures are well-suited for quantifying slow variations of the inputs effects.    
        
\subsubsection{Owen, Dick and Chen's $L^p$ measure of dependence}
For independent random variables and the kernel $k_{l_p} (Y, Y') := |Y|^{2p} |Y'|^{2p}$, we obtain the $L^p$-measure of dependence for the first-order only, which has been proposed in \cite{owen14}, that is, 
$ 
\sqrt{S_{u}^{k_{l_p}}} = \esp \left[\left|\M_u^{fo}(\bo{\X}_u) \right|^{2p}\right]  
$.     
When $p=1$ and inputs are independent, we obtain Sobol' indices. 
            
\subsubsection{The $\ell_2$-based kernel indices: Sobol' indices and generalized sensitivity indices} 
When $\NN =1$, $k_2\left(y, y' \right) := |y|^2|y'|^2$ and the inputs are independent, we can check that the first-order and total Sobol indices are the square root of the $\ell_2$-based kernel SIs, that is,
$$
S_u = \sqrt{S_{u}^{k_2}},  
\qquad \quad       
S_{T_u} := \sqrt{S_{T_u}^{k_2}} \, .      
$$  
Moreover, we have $\sqrt{S_{T_u}^{k_2}} =\frac{1}{2}  \sqrt{\Upsilon_u^{k_2}}$.      

For $\NN >1$, $k_2\left(\bo{y}, \bo{y}' \right) := \norme{\bo{y}}^2 \norme{\bo{y}'}^2$,  we obtain the square of the generalized sensitivity indices (GSIs) of the first-type (see \cite{lamboni11,gamboa14,lamboni18a,lamboni22}). Furthermore, when the target inputs are dependent, the $\ell_2$-based kernel SIs are the square of the dependent GSIs (including GSIs) of the first-type introduced in \cite{lamboni21}, that is, 
$$ 
dGSI^{1,M}_u  =  \sqrt{S_{u}^{k_2}} 
\qquad \quad
dGSI^{1,M}_{T_u} = \sqrt{S_{T_u}^{k_2}} \,  .      
$$  
      
\subsubsection{Quadratic kernel indices and dependent generalized sensitivity indices} 
For $\NN \geq 1$, the kernel $k_q\left(\bo{y}, \bo{y}' \right) := \left<\bo{y}, \, \bo{y}'\right>^2$ leads to  the square of i) GSIs of the second-type for independent variables (see \cite{lamboni11,lamboni18a,lamboni22}), and ii) the dependent GSIs of the second-type (see \cite{lamboni21,lamboni21ar}), that is,   
$$ 
dGSI^{2,M}_u  =  \sqrt{S_{u}^{k_q}} 
\qquad \quad
dGSI^{2,M}_{T_u} = \sqrt{S_{T_u}^{k_q}} \; \leq \; 
\sqrt{\Upsilon_u^{k_p}} \, .
$$  
Thus, $\sqrt{\Upsilon_u^{k_p}}$ becomes the upper bound of the total second-type GSI of $\bo{\X}^w_u$. 
  
\section{Computing the kernel-based sensitivity indices} \label{sec:est}
For computing the first-order and total Kb-SIs in one hand, and the screening measure $\Upsilon_u^k$ in the other hand, we are going to express such indices using the weight function $w_e$ and the CDF $F_{ind}$ (see Proposition \ref{prop:wexpksi}) because the dependency models of $\bo{\X}^w \sim F_{ind}^{w_e}$ have been derived using $w_e$. To that end, we use $\bo{Y}^{'}$, $\bo{Y}^{''}$ and $\bo{Y}^{'''}$ for i.i.d. copies of $\bo{Y} \sim F_{ind}$. 
             
\begin{prop} \label{prop:wexpksi} 
Assume that (A1)-(A2) hold. Then, we have   
\begin{equation}
S_u^{k} := \frac{\esp\left[k\left(\M_u^{fo}(\bo{Y}_u),\, \M_u^{fo}(\bo{Y}^{'}_u) \right) w_e\left(\bo{Y}\right) w_e\left(\bo{Y}'\right) \right]}{\esp\left[k\left(\M^c(\bo{Y}_u, \bo{U}),\, \M^c(\bo{Y}^{'}_u, \bo{U}') w_e\left(\bo{Y}\right) w_e\left(\bo{Y}'\right) \right)\right]} \, ;  
\end{equation}    
\begin{equation}
S_{T_u}^{k} :=  \frac{\esp\left[k\left(\M_u^{tot}(\bo{Y}_u, \bo{U}),\, \M_u^{tot}(\bo{Y}^{'}_u, \bo{U}') \right) w_e\left(\bo{Y}\right) w_e\left(\bo{Y}'\right) \right]}{\esp\left[k\left(\M^c(\bo{Y}_u, \bo{U}),\, \M^c(\bo{Y}^{'}_u, \bo{U}') \right) w_e\left(\bo{Y}\right) w_e\left(\bo{Y}'\right)\right]} \, ; 
\end{equation}   
\begin{equation}   
\Upsilon_u^k :=\frac{\esp\left[k\left(\M^*_u(\bo{Y}_u, \bo{Y}^{'}_u,\bo{U}),\, \M^*_u(\bo{Y}^{''}_u, \bo{Y}^{'''}_u, \bo{U}') \right) w_e\left(\bo{Y}\right) w_e\left(\bo{Y}'\right) w_e\left(\bo{Y}^{''}\right) w_e\left(\bo{Y}^{'''}\right) \right]}{\esp\left[k\left(\M^c(\bo{Y}_u, \bo{U}),\, \M^c(\bo{Y}^{'}_u, \bo{U}') \right) w_e\left(\bo{Y}\right) w_e\left(\bo{Y}'\right)\right] \esp^2\left[w_e\left(\bo{Y}\right)\right]} \,  .  
\end{equation} 
\end{prop}   
\begin{preuve}   
Using Definition \ref{def:kbsi}, the results are straightforward bearing in mind Remark~\ref{rem:mcpdfw}. 
\begin{flushright}  
$\Box$ 
\end{flushright}
\end{preuve}  

To construct the estimators of the Kb-SIs, we are given two independent samples, that is, 
$\left\{\bo{Y}_i, \bo{Y}'_i, \bo{U}_i, \bo{U}'_i\right\}_{i=1}^{m_1}$ and  $\left\{\bo{Y}_i, \bo{Y}'_i, \bo{U}_i, \bo{U}'_i\right\}_{i=1}^{m}$ from $\left(\bo{Y}, \bo{Y}', \bo{U}, \bo{U}'\right)$, and we consider the following consistent estimators: 

$$ 
\widehat{\mu}\left(\bo{Y}_u\right) := \frac{1}{m_1} \sum_{i=1}^{m_1} \M\left(\bo{Y}_u, r\left(\bo{Y}_u, \bo{U}_i \right) \right) 
\; \xrightarrow{P}  \; 
\esp_{\bo{U}}\left[\M\left(\bo{Y}_u, r\left(\bo{Y}_u, \bo{U} \right)\right)  \right] \, ;
$$ 
$$   
\widehat{\mu}\left(\bo{U}\right) := \frac{\sum_{i=1}^{m_1} \M\left(\bo{Y}_{i,u}, r\left(\bo{Y}_{i,u}, \bo{U} \right) \right) w_e\left(\bo{Y}_i \right)}{\sum_{i=1}^{m_1} w_e\left(\bo{Y}_i \right)}   
\; \xrightarrow{P}  \; 
\frac{\esp_{\bo{Y}}\left[\M\left(\bo{Y}_u, r\left(\bo{Y}_u, \bo{U} \right) \right)  w_e(\bo{Y}) \right]}{\esp\left[ w_e(\bo{Y})\right]} \, ,  
$$    
$$  
\widehat{\mu} := \frac{\sum_{i=1}^{m_1} \M\left(\bo{Y}_{i,u}, r\left(\bo{Y}_{i,u}, \bo{U}_i\right) \right) w_e\left(\bo{Y}_i \right)}{\sum_{i=1}^{m_1} w_e\left(\bo{Y}_i \right)}   
\; \xrightarrow{P}  \; 
\frac{\esp\left[\M\left(\bo{Y}_u, r\left(\bo{Y}_u, \bo{U} \right) \right)  w_e(\bo{Y})  \right]}{\esp\left[ w_e(\bo{Y})\right]} \, .
$$
    
Using the method of moments, we derive the estimators of the first-order and total SFs, the centered model outputs as follows: 
$$
\widehat{\M_{u}^{fo}}\left(\bo{Y}_u\right) := \widehat{\mu}(\bo{Y}_{u}) - \widehat{\mu}
\; \xrightarrow{P}  \;  \M_{u}^{fo} \left(\bo{Y}_u\right) \, ; 
$$ 
$$
\widehat{\M_{u}^{tot}}\left(\bo{Y}_u, \bo{U}\right) := \M\left(\bo{Y}_{u}, r\left(\bo{Y}_{u}, \bo{U} \right) \right) - \widehat{\mu}(\bo{U}) \; \xrightarrow{P}  \;  \M_{u}^{tot} \left(\bo{Y}_u, \bo{U}\right) \, ; 
$$ 
$$
\widehat{\M^{c}}\left(\bo{Y}_u, \bo{U}\right) := \M\left(\bo{Y}_{u}, r\left(\bo{Y}_{u}, \bo{U} \right) \right) - \widehat{\mu} \; \xrightarrow{P}  \;  \M^{c} \left(\bo{Y}_u, \bo{U}\right) \, . 
$$ 
	
Again the method of moments and the plug-in approach allow for deriving the consistent estimators of the Kb-SIs in Theorem \ref{theo:estksi}.  To that end, we use 
$$
\sigma^{fo}_u := \var\left[k\left(\M_{u}^{fo}\left(\bo{Y}_{u}\right),\, \M_{u}^{fo}\left(\bo{Y}_{u}'\right) \right) w_e\left(\bo{Y} \right) w_e\left(\bo{Y}' \right) \right] \, ,
$$  
$$ 
\sigma^{tot}_u := \var\left[
k\left(\M^{tot}_u\left(\bo{Y}_{u}, \bo{U} \right),\, \M^{tot}_u\left(\bo{Y}_{u}', \bo{U}' \right) \right) w_e\left(\bo{Y} \right) w_e\left(\bo{Y}' \right)\right] \, , 
$$
$$
\mu_c^k := \esp\left[k\left(\M^{c}\left(\bo{Y}_{u}, \bo{U} \right),\, \M^{c}\left(\bo{Y}_{u}', \bo{U}' \right) \right) w_e\left(\bo{Y} \right) w_e\left(\bo{Y}' \right) \right] \, . 
$$
To derive the asymptotic distributions of the estimators of Kb-SIs, we suppose that we use a sample of size $M\gg m$ to estimate $\mu_c^k$. Indeed, for estimating the non-normalized Kb-SIs of $\bo{\X}_u$ for all $u\subseteq \{1, \ldots, d\}$, different samples are going to be used, and some of such samples can be combined for estimating $\mu_c^k$. The consistent estimator of $\mu_c^k$ is given by
$$
\widehat{\mu_c^k} := \frac{1}{M} \sum_{i=1}^M k\left(\widehat{\M^{c}}\left(\bo{Y}_{i,u}, \bo{U}_i \right),\, \widehat{\M^{c}}\left(\bo{Y}_{i,u}', \bo{U}_i' \right) \right) w_e\left(\bo{Y}_{i} \right) w_e\left(\bo{Y}_{i}' \right) \; \xrightarrow{P}  \;   \mu_c^k \, .    
$$
             
\begin{theorem} \label{theo:estksi}    
Assume that (A1)-(A2) hold and $k$ is differentiable almost everywhere. 
 
$\quad$ (i) The consistent estimators of $S_{u}^k$ and $S_{T_u}^k$ are respectively given by 
\begin{equation}  \label{eq:estksifo}  
\widehat{S_{u}^k} :=   \frac{\frac{1}{m} \sum_{i=1}^m k\left(\widehat{\M_{u}^{fo}}\left(\bo{Y}_{i,u}\right),\, \widehat{\M_{u}^{fo}}\left(\bo{Y}_{i,u}'\right) \right) w_e\left(\bo{Y}_{i} \right) w_e\left(\bo{Y}_{i}' \right)}{\frac{1}{M} \sum_{i=1}^M k\left(\widehat{\M^{c}}\left(\bo{Y}_{i,u}, \bo{U}_i \right),\, \widehat{\M^{c}}\left(\bo{Y}_{i,u}', \bo{U}_i' \right) \right) w_e\left(\bo{Y}_{i} \right) w_e\left(\bo{Y}_{i}' \right)} \, ; 
\end{equation}      
\begin{equation} \label{eq:estksitot}  
\widehat{S_{T_u}^k} := \frac{\frac{1}{m} \sum_{i=1}^m k\left(\widehat{\M^{tot}_u}\left(\bo{Y}_{i,u}, \bo{U}_i \right),\, \widehat{\M^{tot}_u}\left(\bo{Y}_{i,u}', \bo{U}_i' \right) \right) w_e\left(\bo{Y}_{i} \right) w_e\left(\bo{Y}_{i}' \right)}{ \frac{1}{M} \sum_{i=1}^M k\left(\widehat{\M^{c}}\left(\bo{Y}_{i,u}, \bo{U}_i \right),\, \widehat{\M^{c}}\left(\bo{Y}_{i,u}', \bo{U}_i' \right) \right) w_e\left(\bo{Y}_{i} \right) w_e\left(\bo{Y}_{i}' \right)}  \, .       
\end{equation}
    
$\quad$ (ii) If $m_1, m, M \to \infty$ with $\frac{m}{M} \to 0;\, \frac{m_1}{M} \to 0$, then   
$$    
\sqrt{m}\left(\widehat{S_{u}^k} - S_{u}^k \right) \, \xrightarrow{D} \, \mathcal{N} \left(0, \frac{\sigma^{fo}_u}{\left(\mu_c^k\right)^2} \right);
\qquad \quad 
\sqrt{m}\left(\widehat{S_{T_u}^k} - S_{T_u}^k \right) \, \xrightarrow{D} \,  \mathcal{N} \left(0, \frac{\sigma^{tot}_u}{\left(\mu_c^k\right)^2} \right) \, . 
$$    
\end{theorem}  
\begin{preuve}
See Appendix \ref{app:theo:estksi}.   
\begin{flushright}
$\Box$ 
\end{flushright} 
\end{preuve}

\section{Test cases} \label{sec:test}            
We consider two functions to illustrate our approach. The first test case treats a cluster of the output values by making use of the CDF $W$ (see Theorem \ref{theo:copbcd}) to derive analytical target input distributions and SFs. The second test case deals with a vector-valued function and a polynomial weight function. To compute the kernel-based SIs, we consider the $\ell_1$-based kernel and the quadratic kernel.  
         
\subsection{Quadratic function ($d=3$, $\NN=1$)} \label{sec:test1}   
In  this section, we consider the following single response model: 
\begin{equation} \label{eq:line}        
    \M(\bo{\X}) = \X_1^2 + \X_2^2 +  \X_3^2  \, ,       \nonumber   
\end{equation} 
which includes three independent input factors $\X_j \sim \mathcal{N}\left(0, \, 1 \right)$ with $j=1, 2, 3$. For a given threshold $c \in \R_+$, we are interested in the target input distribution given by 
$$
dF^{w}(\bo{\x}) = \frac{\indic_{]-\infty, \, c]}(\M(\bo{x}))}{\esp_{F_{ind}}\left[\indic_{]-\infty, \, c]}(\M(\bo{\X})) \right]} \, dF_{ind}(\bo{\x}); \quad \forall\;  \bo{x} \in \R^d\, . 
$$   
Such model behavior represents a safe domain such as the concentration of a biological hazard below a given threshold $c$, that is, $\M(\bo{\X}) \leq c$. To derive the analytical SFs, it is known that a dependency model of    
$$ 
\bo{\X}^w :=\left(\X_1^w,  \X_2^w,  \X_3^w \right) :=
\left\{\X_j \sim \mathcal{N}(0, 1), j=1, 2, 3 : \X_1^2 +\X_2^2 +\X_3^2 \leq c \right\}
$$  
is given by    
$  
(\X_2^w)^2 = Z_2(c-(\X_1^w)^2),\quad  (\X_3^w)^2 = Z_3(c-(\X_1^w)^2)(1-Z_2) 
$,      
where $(\X_1^w)^2 \sim B1(c, 1/2, 2)$,  $Z_2 \sim Beta(1/2, 3/2)$ and $Z_3 \sim Beta(1/2, 1)$ are independent with $B1$ the beta distribution of first-kind (see \cite{lamboni22b}, Corollaries 2, 4 for more details). Thus, we have (see  Appendix \ref{app:sfs})
$$
\M(\bo{\X}^w) = (\X_1^w)^2(1 -Z_2 - Z_3(1-Z_2)) + cZ_2 +  cZ_3(1-Z_2) \, ; 
$$      
$$
\M^{c}(\X_1^w, Z_2, Z_3) = (\X_1^w)^2(1 -Z_2 - Z_3(1-Z_2)) + cZ_2 +  cZ_3(1-Z_2) - \frac{3}{5} c \, . 
$$   
Moreover, the first-order and total SFs of $\X_1^w$ and $\M^*$ are given by  
$$
\M_1^{fo}(\X_1^w) = \frac{1}{2} \left[(\X_1^w)^2 - c/5 \right]; 
\qquad
\M_1^{tot}(\X_1^w, Z_2, Z_3) = \left[(\X_1^w)^2 - c/5 \right] (1 -Z_2)(1-Z_3)
\, ;  
$$   
$$ 
\M^{*}_1(\X_1^w, \X_1^{w'}, Z_2, Z_3) = \left[(\X_1^w)^2- \left(\X_1^{w'} \right)^2 \right](1 -Z_2)(1-Z_3) \, .    
$$

The computations of Kb-SIs using the above functionals give the following results. For the $\ell_1$-based kernel, the indices of $\X_j^w$ are $\sqrt{S_{j}^{k_1}} = \sqrt{S_{T_j}^{k_1}}=0.385$ and $\sqrt{\Upsilon_j^{k_1}} =0.505$ with $j=1, 2, 3$. For the quadratic kernel, we have obtained $\sqrt{S_{j}^{k_q}} =0.167$ $\sqrt{ S_{T_j}^{k_q}}=0.223$ and $\sqrt{\Upsilon_j^{k_q}} =0.445$ with $j=1, 2, 3$. These indices are invariant by changing the threshold $c$.  
    
\subsection{Multivariate g-Sobol function ($d=10$, $\NN=4$)} \label{sec:test3}   
We consider the multivariate g-Sobol function, which includes ten independent input variables following the standard uniform distribution (i.e., $\X_j \sim \mathcal{U}(0,\, 1)$, $j=1, \ldots, 10$) and provides four outputs. It is given by  (\cite{lamboni18a,lamboni22})
$$  
\small{ 
\M(\bo{\x}) :=  \left[ \begin{array}{c} 
\prod_{j=1}^{d=10}\frac{ |4\, \x_j \,- \,2| \,+ \,\mathcal{A}[1,j]}{1 \,+\, \mathcal{A}[1,j]} \\ 
\prod_{j=1}^{d=10}\frac{ |4\, \x_j \,-\, 2| \,+ \,\mathcal{A}[2,j]}{1 \,+ \,\mathcal{A}[2,j]} \\ 
\prod_{j=1}^{d=10}\frac{ |4 \, \x_j \,- \,2|\, + \,\mathcal{A}[3,j]}{1 \,+\, \mathcal{A}[3,j]} \\ 
\prod_{j=1}^{d=10}\frac{ |4 \, \x_j \,- \,2| \, + \,\mathcal{A}[4,j]}{1 \,+\, \mathcal{A}[4,j]} 
\end{array}      
\right],     
\, 
\mathcal{A} =\left[ \begin{array}{cccccccccc}
0 & 0 & 6.52 & 6.52 & 6.52 & 6.52 & 6.52 & 6.52 & 6.52 & 6.52 \\
0 & 1 & 4.5 & 9 & 99 & 99 & 99 & 99 & 99 & 99 \\ 
1 & 2 & 3 & 4 & 5 & 6 & 7 & 8 & 9 & 10 \\ 
50 & 50 & 50 & 50 & 50 & 50 & 50 & 50 & 50 & 50 
\end{array}  \right]} \, . 
$$                  
We are interested in the model behavior defined by $w(\bo{x}) := \prod_{j=1}^{10} x_j^{\alpha_j}$ for every $\alpha_j \in \R_+$. We can see that the target inputs distribution is given by 
$\rho^w(\bo{x}) = \frac{\prod_{j=1}^{10} x_j^{\alpha_j} \indic_{[0, \, 1]}(x_j)}{\esp_{F_{ind}}\left[\prod_{j=1}^{10} \X_j^{\alpha_j}\right]}$, which implies that $\X_j^w \sim Beta(\alpha_j+1, 1)$ for any $j \in \{1, \ldots, d\}$. \\
Tables \ref{tab:sobA1}-\ref{tab:sobA50} provide the  estimates of sensitivity indices using the $\ell_1$-based kernel and the quadratic kernel. Obviously, when $q=1/2$ and $\boldsymbol{\alpha}=(0, \ldots, 0)$ (see Table \ref{tab:sobA1}), we obtain the estimations of GSIs provided in \cite{lamboni22} for the quadratic kernel. Only $\X_1$ and $\X_2$ are important according to the total GSIs and their upper bounds. Similar results were obtained using the $\ell_1$-based kernel SIs if we fix the threshold at $T=0.2$. Decreasing the threshold to $0.1$ shows that two more inputs, that is, $\X_3,\, \X_4$ are important.  This difference is probably due to the fact that the $\ell_1$-based kernel measures slow variations of SFs compared to the quadratic kernel, which is associated with the $L_2$-norm for single valued functions.      
\begin{table}[ht]      
\centering    
\begin{tabular}{lcccccccccc} 
  \hline
	  \hline    
Kernels		 & $\X_1$ & $\X_2$ & $\X_3$ & $\X_4$  & $\X_5$ & $\X_6$ & $\X_7$ & $\X_8$ & $\X_9$ & $\X_{10}$ \\   
 \hline   
		\multicolumn{11}{c}{First-order Kb-SIs} \\ 
$\ell_1$-based & 0.756 & 0.556 & 0.176 & 0.136 & 0.099 & 0.092 & 0.086 & 0.082 & 0.079 & 0.076 \\ 
Quadratic & 0.536 & 0.328 & 0.027 & 0.016 & 0.011 & 0.009 & 0.008 & 0.007 & 0.007 & 0.006 \\
  \hline 
		\multicolumn{11}{c}{Total Kb-SIs} \\
$\ell_1$-based & 0.756 & 0.555 & 0.178 & 0.148 & 0.099 & 0.091 & 0.085 & 0.082 & 0.078 & 0.076 \\ 
Quadratic & 0.637 & 0.436 & 0.038 & 0.026 & 0.015 & 0.012 & 0.011 & 0.011 & 0.009 & 0.010 \\ 
  \hline  
		\multicolumn{11}{c}{Upper bounds of Kb-SIs} \\
$\ell_1$-based & 1.002 & 0.747 & 0.235 & 0.188 & 0.133 & 0.124 & 0.114 & 0.114 & 0.105 & 0.100 \\ 
Quadratic & 1.230 & 0.844 & 0.074 & 0.048 & 0.031 & 0.027 & 0.021 & 0.024 & 0.020 & 0.018 \\ 
	\hline        
	\hline  
\end{tabular} 
\caption{Square root of kernel-based SIs for the multivariate Sobol function associated with $\boldsymbol{\alpha}=(0, \ldots, 0)$.}
\label{tab:sobA1} 
\end{table}  

By taking $\boldsymbol{\alpha}=(20, 20, 10, 10, 10, 10, 10, 1, 1, 1)$, the $\ell_1$-based kernel SIs show that all the inputs are important ($T=0.1$) while the quadratic kernel SIs identify $\X_j^w, \, j=3, \ldots, 7$ as non important variables (see Table \ref{tab:sobA50}). As expected, the results provided in Tables \ref{tab:sobA1}-\ref{tab:sobA50} prove that changing the model behavior of interest (thanks to the weight functions) can significantly modify the impacts of the input variables. The polynomial weight function $x_j^{\alpha_j}$ aims at emphasizing the selection of the initial values of $\X_j$ that are close to $1$ when the real $\alpha_j$ is large. Thus,  it restricts the support of $\X_j$, which results in decreasing the effect of $\X_j^w$ on the model outputs.   
\begin{table}[ht]            
\centering   
\begin{tabular}{lcccccccccc} 
  \hline
	  \hline  
Kernels & $\X_1$ & $\X_2$ & $\X_3$ & $\X_4$  & $\X_5$ & $\X_6$ & $\X_7$ & $\X_8$ & $\X_9$ & $\X_{10}$ \\     
 \hline 
		\multicolumn{11}{c}{First-order Kb-SIs} \\ 
$\ell_1$-based & 0.465 & 0.388 & 0.277 & 0.229 & 0.173 & 0.165 & 0.157 & 0.319 & 0.309 & 0.301 \\ 
Quadratic & 0.261 & 0.190 & 0.075 & 0.058 & 0.041 & 0.040 & 0.035 & 0.121 & 0.116 & 0.113 \\ 
  \hline 
		\multicolumn{11}{c}{Total Kb-SIs} \\
$\ell_1$-based & 0.465 & 0.388 & 0.276 & 0.229 & 0.173 & 0.164 & 0.157 & 0.320 & 0.309 & 0.300 \\ 
Quadratic & 0.267 & 0.197 & 0.074 & 0.059 & 0.043 & 0.041 & 0.037 & 0.123 & 0.120 & 0.116 \\  
  \hline  
		\multicolumn{11}{c}{Upper bounds of Kb-SIs} \\
$\ell_1$-based & 0.615 & 0.530 & 0.374 & 0.303 & 0.238 & 0.224 & 0.211 & 0.428 & 0.417 & 0.402 \\ 
Quadratic & 0.477 & 0.441 & 0.148 & 0.104 & 0.088 & 0.081 & 0.069 & 0.252 & 0.252 & 0.233 \\ 
 \hline         
	\hline 
\end{tabular}   
\caption{Square root of kernel-based SIs for the multivariate Sobol function associated with $\boldsymbol{\alpha}=(20, 20, 10, 10, 10, 10, 10, 1, 1, 1)$.}  
\label{tab:sobA50} 
\end{table}       
                                                      
\section{Conclusion} \label{sec:con}   
We have proposed a methodology for exploring specific model behaviors such as identifying the input variables that drive the initial model output(s) toward a domain of interest (e.g., failure, safe and sustainable domain) and/or govern the target output(s) defined via weight functions, including the output values within a given cluster. Weight functions include rule based ensembles, non-negative desirability measures of the model output(s) to meet a given criterion, membership functions from crisp or fuzzy clustering, and any non-negative function based on classifiers such as PCA, kernel PCA, logistic regression, random forest. The proposed approach is well-suited for performing SA not only for the target output(s) but also for  models with inputs following the multivariate weighted distribution, including the copula-based distributions.    
      
We have introduced the kernel-based SIs, including the $\ell_1$-based SIs that aim at addressing the issues of high-order moments of SFs and interactions among the model drivers even non-independent. The well-kown variance-based SA and dependent multivariate SA are particular cases of our new measures of association between model drivers and outputs. We have provided consistent estimators of kernel-based SIs for computing such indices by distinguishing the case of multivariate and functional outputs (including spatio-temporal and dynamic models) and the case of  multivariate response models, including single response models.
                  
Analytical and numerical results show the relevance of our approach for analyzing a given model behavior. For the second test case, our results are equal to those provided in \cite{lamboni22,lamboni21,lamboni21ar}, emphasizing the extension of generalized sensitivity indices and dependent generalized sensitivity indices. In next future, it is worth investigating the extension of this work to cope with most kernels and the original space of labels used as outputs.  
  
                
 \section*{Acknowledgments}
 We would like to thank S. Buis and S. Roux for some useful discussions.             
     
\begin{appendices}
\section{Proof of Proposition \ref{prop:neww}} \label{app:prop:neww}
Knowing that the density of $\bo{\X}$ is given by 
$
\rho(\bo{x}) = c\left(F_1(x_1), \ldots, F_{d}(x_{d})\right) \prod_{j=1}^d  \rho_j(x_j)
$, 
we can write thanks to Equation (\ref{eq:wdist}) 
$$
\rho^w(\bo{\x}) =  \frac{w(\bo{x})}{\esp_F\left[w(\bo{\X}) \right]}
\, c\left(F_1(x_1), \ldots, F_{d}(x_{d})\right) \prod_{j=1}^d  \rho_j(x_j) \, . 
$$   
         
\section{Proof of Theorem \ref{theo:copbcd}} \label{app:theo:copbcd}
The following proof is simpler than the general one provided in \cite{lamboni22mcap}.
Since $\rho^w(\bo{\x}) =  \frac{w_e(\bo{x})}{\esp_{F_{ind}}\left[w_e(\bo{Y}) \right]} \prod_{j=1}^d  \rho_j(x_j)$ (see Proposition \ref{prop:neww}), the density of $\bo{\X}_{\sim u}^w | \bo{\X}_{u}^w$ becomes 
\begin{equation} \label{eq:wcdfind}
\rho_{\sim u|u}^{w}(\bo{x}_{\sim u} | \bo{x}_{u}^w) = \frac{w_e(\bo{x}^w_u, \bo{x}_{\sim u})}{\esp_{F_{ind}} \left[w_e(\bo{x}^w_u, \bo{Y}_{\sim u}) \right]} \prod_{j \in (\sim u)} \rho_j(x_j)  \, ,  \nonumber  
\end{equation}  
and we can write  (bearing in mind $(\sim u)=(\pi_1, \dots, \pi_{|\boldsymbol{\pi}|})$)
  
\begin{equation} \label{eq:wcdfind2}   
F_{\sim u|u}^{w}(\bo{x}_{\sim u} | \bo{x}_{u}^w) =  = \esp_{F_{ind}} \left[\frac{w_e(\bo{x}^w_u, \bo{Y}_{\sim u})}{\esp_{F_{ind}} \left[w_e(\bo{x}^w_u, \bo{Y}_{\sim u}) \right]} \prod_{j=1}^{|\boldsymbol{\pi}|} \indic_{[-\infty,\, x_{\pi_j} ]}(Y_{\pi_j})
\right]  \nonumber \, .   
\end{equation}        
Knowing that $\X_{\pi_j} \stackrel{d}{=} F_{\pi_j}^{-1}(U_{\pi_j})$ with $U_{\pi_j} \sim \mathcal{U}(0, \,1)$ and using the theorem of transfer, we have    
\begin{equation} 
F_{\sim u|u}^{w}(\bo{x}_{\sim u} | \bo{x}_{u}^w) = \esp_{\bo{U}_{\boldsymbol{\pi}}} \left[\frac{w_{e}\left(\bo{x}^w_u, F_{\pi_1}^{-1}(U_{\pi_1}), \ldots, F_{\pi_{|\boldsymbol{\pi}|}}^{-1}(U_{\pi_{|\boldsymbol{\pi}|}}) \right)}{\esp_{F_{ind}} \left[w_e(\bo{x}^w_u, \bo{Y}_{\sim u}) \right]} \prod_{j=1}^{|\boldsymbol{\pi}|} \indic_{[-\infty,\, x_{\pi_j} ]}\left(F_{\pi_j}^{-1}(U_{\pi_j}) \right) \right]  \nonumber \, ,
\end{equation}       
with $\bo{U}_{\boldsymbol{\pi}} := \left(U_{\pi_j},\, j=1,\ldots, |\boldsymbol{\pi}| \right) \sim \mathcal{U}\left(0, \, 1\right)^d$.
As  $F_j$ is strictly increasing, we have  
\begin{eqnarray}         
F_{\sim u|u}^{w}(\bo{x}_{\sim u} | \bo{x}_{u}^w) &=& \esp_{\bo{U}_{\boldsymbol{\pi}}} \left[\frac{w_{e}\left(\bo{x}^w_u, F_{\pi_1}^{-1}(U_{\pi_1}), \ldots, F_{\pi_{|\boldsymbol{\pi}|}}^{-1}(U_{\pi_{|\boldsymbol{\pi}|}}) \right)}{\esp_{F_{ind}} \left[w_e(\bo{x}^w_u, \bo{Y}_{\sim u}) \right]} \prod_{j=1}^{|\boldsymbol{\pi}|} \indic_{[0,\, F_{\pi_j}(x_{\pi_j}) ]} \left(U_{\pi_j} \right) \right]  \nonumber \\
&=&   
\int_{0}^{F_{\pi_1}(x_{\pi_1})}\ldots  \int_{0}^{F_{\pi_{|\boldsymbol{\pi}|}}(x_{\pi_{|\boldsymbol{\pi}|}})} 
 \frac{w_{e}\left(\bo{x}^w_u, F_{\pi_1}^{-1}(v_{\pi_1}), \ldots, F_{\pi_{|\boldsymbol{\pi}|}}^{-1}(v_{\pi_{|\boldsymbol{\pi}|}}) \right)}{\esp_{F_{ind}}\left[w_{e}(\bo{x}^w_u, \bo{Y}_{\sim u}) \right]}  \prod_{j=1}^{|\boldsymbol{\pi}|} dv_{\pi_j} \,  . \nonumber 
\end{eqnarray}
        
Now, if we use $\bo{V}_{\boldsymbol{\pi}} :=\left(V_{\pi_k} \sim \mathcal{U}(0,\, u_{\pi_k}), \, k=1, \ldots, |\boldsymbol{\pi}| \right)$ for a random vector of independent variables and 
\begin{eqnarray}    
W\left(\bo{u}_{\boldsymbol{\pi}}; \bo{x}_{u}^w \right) 
&:=& \int_{0}^{u_{\pi_1}}\ldots  \int_{0}^{u_{\pi_{|\boldsymbol{\pi}|}}} 
 \frac{w_{e}\left(\bo{x}^w_u, F_{\pi_1}^{-1}(v_{\pi_1}), \ldots, F_{\pi_{|\boldsymbol{\pi}|}}^{-1}(v_{\pi_{|\boldsymbol{\pi}|}}) \right)}{\esp_{F_{ind}}\left[w_{e}(\bo{x}^w_u, \bo{Y}_{\sim u}) \right]}  \prod_{j=1}^{|\boldsymbol{\pi}|} dv_{\pi_j} \nonumber  \\
&=& \frac{\esp_{\bo{V}_{\boldsymbol{\pi}}} \left[w_{e}\left(\bo{x}^w_u, F_{\pi_1}^{-1}(V_{\pi_1}), \ldots, F_{\pi_{|\boldsymbol{\pi}|}}^{-1}(V_{\pi_{|\boldsymbol{\pi}|}}) \right) \right]}{\esp_{F_{ind}}\left[w_{e}(\bo{x}^w_u, \bo{Y}_{\sim u}) \right]}\,  \prod_{j=1}^{|\boldsymbol{\pi}|} u_{\pi_j} \nonumber\, ,
\end{eqnarray}        
then $W$ is a CDF of a random vector having $(0, \, 1)^{d-|u|}$ as the support, and  we have      
$$ 
F_{\sim u|u}^{w}(\bo{x}_{\sim u} | \bo{x}_{u}^w) = W\left(F_{\pi_1}(x_{\pi_1}), \ldots, F_{\pi_{|\boldsymbol{\pi}|}}(x_{\pi_{|\boldsymbol{\pi}|}}); \bo{x}_{u}^w \right) \, .  
$$   

\section{Proof of Lemma \ref{lem:pro}} \label{app:lem:pro}
Since $\bo{\X}^{w}_u$, $\bo{\X}^{w '}_u$ are i.i.d., we can write thanks to Proposition~\ref{prop:tef}   
\begin{eqnarray}  
\M_u^{fo}(\bo{\X}^w_u) &=&  \esp_{U} \left[\M\left(\bo{\X}^w_u,  r\left(\bo{\X}_{u}^w, \bo{U} \right)\right)\right] -  
\frac{\esp\left[\M\left(\bo{Y}_u,  r\left(\bo{Y}_{u}, \bo{U} \right)\right) w_e\left(\bo{Y}\right)\right]}{\esp\left[w_e\left(\bo{Y}\right)\right]} \nonumber \\
&=&  \esp_{U} \left[\M\left(\bo{\X}^w_u,  r\left(\bo{\X}_{u}^w, \bo{U} \right)\right) -  
\frac{\esp_{\bo{Y}}\left[\M\left(\bo{Y}_u,  r\left(\bo{Y}_{u}, \bo{U} \right)\right) w_e\left(\bo{Y}\right)\right]}{\esp\left[w_e\left(\bo{Y}\right)\right]} \right] \nonumber \\
&=& \esp_{U} \left[\M_u^{tot}\left(\bo{\X}^w_u, \bo{U} \right) \right] \, . \nonumber 
\end{eqnarray}      
Using the convexity of $\phi$ and the definition of the kernel, the Jensen inequality yields 
$$  
\esp\left[k\left(\M_u^{fo}(\bo{\X}^w_u),\, \M_u^{fo}(\bo{\X}^{w'}_u)\right)\right] \leq 
\esp\left[k\left(\M_u^{tot}\left(\bo{\X}^w_u, \bo{U} \right),\, \M_u^{tot}\left(\bo{\X}^{w'}_u, \bo{U}' \right)\right)\right] \, .  
$$        
  
Now, we are going to show that the total index is less than one. Let us consider the Dirac probability measure $\delta_{\bo{U}}(\bo{U'}) :=\delta_{\bo{0}}(\bo{U'}-\bo{U})$ and the zero-mean expression of the outputs, that is,   \\  
$      
\M_{u}^{c}(\bo{\X}^w_u, \bo{U}) = \esp \left[\M\left(\bo{\X}^w_u,  r\left(\bo{\X}_{u}^w, \bo{U} \right)\right) -  \frac{\esp_{\bo{Y}}\left[\M\left(\bo{Y}_u,  r\left(\bo{Y}_{u}, \bo{U}' \right)\right) w_e\left(\bo{Y}\right)\right]}{\esp\left[w_e\left(\bo{Y}\right)\right]} \, | \bo{U}, \bo{\X}^w_u \right]    
$.  We can then write                    
\begin{eqnarray}          
& & \esp\left[\M_{u}^{c}(\bo{\X}^w_u, \bo{U}) \, | \delta_{\bo{U}}(\bo{U'}), \bo{U}, \bo{\X}^w_u\right] \nonumber \\  
&=& \esp\left[\esp \left[\M\left(\bo{\X}^w_u,  r\left(\bo{\X}_{u}^w, \bo{U} \right)\right) -  \frac{\esp_{\bo{Y}}\left[\M\left(\bo{Y}_u,  r\left(\bo{Y}_{u}, \bo{U}' \right)\right) w_e\left(\bo{Y}\right)\right]}{\esp\left[w_e\left(\bo{Y}\right)\right]} \, | \bo{U}, \bo{\X}^w_u \right]\, | \delta_{\bo{U}}(\bo{U'}), \bo{U}, \bo{\X}^w_u \right] \nonumber \\
&=&\esp\left[ \esp \left[\M\left(\bo{\X}^w_u,  r\left(\bo{\X}_{u}^w, \bo{U} \right)\right) -  \frac{\esp_{\bo{Y}}\left[\M\left(\bo{Y}_u,  r\left(\bo{Y}_{u}, \bo{U}' \right)\right) w_e\left(\bo{Y}\right)\right]}{\esp\left[w_e\left(\bo{Y}\right)\right]} \, | \delta_{\bo{U}}(\bo{U'}), \bo{U}, \bo{\X}^w_u \right] \, | \bo{U}, \bo{\X}^w_u \right] \nonumber \\   
&=& \esp \left[\M\left(\bo{\X}^w_u,  r\left(\bo{\X}_{u}^w, \bo{U} \right)\right) -  \frac{\esp_{\bo{Y}}\left[\M\left(\bo{Y}_u,  r\left(\bo{Y}_{u}, \bo{U} \right)\right) w_e\left(\bo{Y}\right)\right]}{\esp\left[w_e\left(\bo{Y}\right)\right]} \, | \bo{U}, \bo{\X}^w_u \right] =\M_{u}^{tot}(\bo{\X}^w_u, \bo{U})  \nonumber \, ,  
\end{eqnarray}     
bearing in mind the formal definition of conditional expectation.   
The second result holds by applying the conditional Jensen inequality. \\
For the upper bound of the total index, knowing that 
$ 
\M_u^{*}(\bo{\X}^w_u, \bo{\X}^{w'}_u, \bo{U}) = \M\left(\bo{\X}^w_u,  r\left(\bo{\X}_{u}^w, \bo{U} \right)\right) - \M\left(\bo{\X}^{w'}_u,  r\left(\bo{\X}^{w'}_u, \bo{U} \right)\right)$, we can write 
$
\M_u^{tot}(\bo{\X}^w_u, \bo{U}) = \esp_{\bo{\X}^{w'}_u} \left[\M_u^{*}(\bo{\X}^w_u, \bo{\X}^{w'}_u, \bo{U}) \right] \, ,
$  
and the result follows.    

\section{Proof of Theorem \ref{theo:estksi}} \label{app:theo:estksi}
First, the consistency of the estimators holds by applying the Slutsky theorem bearing in mind the Taylor expansion, that is,   
\begin{eqnarray} 
& & k\left(\widehat{\M_{u}^{fo}}\left(\bo{Y}_{i,u}\right),\, \widehat{\M_{u}^{fo}}\left(\bo{Y}_{i,u}'\right) \right)
 = k\left(\M_{u}^{fo}\left(\bo{Y}_{i,u}\right),\, \M_{u}^{fo}\left(\bo{Y}_{i,u}'\right) \right) \nonumber \\ 
 & &  + 
\nabla^\T k\left(\M_{u}^{fo}\left(\bo{Y}_{u}\right),\, \M_{u}^{fo}\left(\bo{Y}_{u}'\right) \right) \left[
\begin{array}{c}
\widehat{\M_{u}^{fo}}\left(\bo{Y}_{i,u}\right) - \M_{u}^{fo}\left(\bo{Y}_{i,u}\right)\\
\widehat{\M_{u}^{fo}}\left(\bo{Y}_{i,u}' \right) - \M_{u}^{fo}\left(\bo{Y}_{i,u}' \right)
\end{array}
 \right]  + R_{m_1}   \, ,  \nonumber 
\end{eqnarray} 
with $R_{m_1}  \xrightarrow{P} 0$ when $m_1 \to \infty$. We obtain the results by applying the law of large numbers\\
Second, the central limit theorem ensures that
$$
\sqrt{m}\left(\frac{1}{m} \sum_{i=1}^m k\left(\widehat{\M_{u}^{fo}}\left(\bo{Y}_{i,u}\right),\, \widehat{\M_{u}^{fo}}\left(\bo{Y}_{i,u}'\right) \right) w_e\left(\bo{Y}_{i} \right) w_e\left(\bo{Y}_{i}' \right) - D_{u}^k \right) \, \xrightarrow{D} \, \mathcal{N} \left(0, \sigma^{fo}_u \right) \, ,  
$$
with  
$
D_{u}^k = \esp\left[k\left(\M_{u}^{fo}\left(\bo{Y}_{u}\right),\, \M_{u}^{fo}\left(\bo{Y}_{u}'\right) \right) w_e\left(\bo{Y} \right) w_e\left(\bo{Y}' \right) \right]
$. \\ 
Third, the asymptotic distributions are straightforward using the Slutsky theorem under the technical assumption $m/M \to 0,\quad  m_1/M \to 0$ (see \cite{lamboni18,lamboni18a} for more details).   
  
\section{Derivation of SFs used  in Section \ref{sec:test1}} \label{app:sfs} 
Using the model output and the dependency models, we can write 
\begin{eqnarray}  
\M_1^{fo}(\X_1^w) &=& (\X_1^w)^2 (1 -\esp[Z_2] - \esp[Z_3(1-Z_2)]) - 
\esp[(\X_1^w)^2](1 -\esp[Z_2] - \esp[Z_3(1-Z_2)]) \nonumber \\
&=& \left[(\X_1^w)^2 - \esp[(\X_1^w)^2] \right] (1 -\esp[Z_2] - \esp[Z_3(1-Z_2)]) \nonumber \\
&=& \left[(\X_1^w)^2 - c/5 \right] (1 -1/4 - 1/3(1-1/4)]) = \frac{1}{2} \left[(\X_1^w)^2 - c/5 \right] \, ; \nonumber 
\end{eqnarray}    
$$       
\M_1^{tot}(\X_1^w, Z_2, Z_3) = \left[(\X_1^w)^2 - \esp[(\X_1^w)^2] \right] (1 -Z_2 - Z_3(1-Z_2)) = \left[(\X_1^w)^2 - c/5 \right] (1 -Z_2 - Z_3(1-Z_2)) \, . 
$$        
We also have        
$$
\M^{c}(\X_1^w, Z_2, Z_3) = \M(\bo{\X}^w) - c/10 -c/4 -c/4 =
 (\X_1^w)^2(1 -Z_2 - Z_3(1-Z_2)) + cZ_2 +  cZ_3(1-Z_2) - \frac{3}{5} c \, . 
$$ 
$$
\M^{*}_1 (\X_1^w, \X_1^{w'}, Z_2, Z_3) =  \left[(\X_1^w)^2- \left(\X_1^{w'} \right)^2 \right](1 -Z_2 - Z_3(1-Z_2)) \, . 
$$

\end{appendices}  
 

\end{document}